\documentstyle[12pt]{article}
\begin{document}
\def\th{\theta}
\def\de{\Delta}
\def\cj{{\Im}}
\def\kb{k_\beta}
\def\beq{\begin{equation}}
\def\enq{\end{equation}}
\def\beqn{\begin{eqnarray}}
\def\eenq{\end{eqnarray}}
\def\pl{\parallel}

\begin{flushright}
MRI-PHY/96/03
\end{flushright}

\begin{center}
{\large{\bf Fluctuation Effects And Order Parameter Symmetry  
    In The Cuprate Superconductors }}\\
\vspace{0.8cm} 
{\bf Biplab Chattopadhyay, D. M. Gaitonde and A. Taraphder}\\

{Mehta Research Institute\\ 
10 Kasturba Gandhi Marg, Allahabad 211002, India.} 
\end{center} 
\vspace{1cm}

\begin{abstract}

Effect of phase fluctuations on superconducting states with anisotropic 
order parameters is studied in a BCS like lattice model of cuprate  
superconductors. The degradation of the mean field transition temperature 
due to phase fluctuations is estimated within a Kosterlitz-Thouless 
scenario. Values of the interaction parameters for optimal doping, 
corresponding to a stable superconducting state of $S_{xy}$ symmetry,  
which fit the nodal structure of the superconducting order parameter in 
the Bi2212 compound, are obtained. The angular 
position of the node is found to be 
insensitive to the dopant concentration.

\end{abstract}
\vspace{1ex} 

\newpage 

The symmetry of the order parameter (OP)  in high temperature 
superconductors continues to be a matter of much controversy. 
A series of experiments [1-6] have indicated that the  
OP in high temperature superconductors is highly 
anisotropic \cite{rmp}. 
However, these experiments have conflicting features [8,9]
 which have not yet made it possible to determine the 
symmetry of the OP in an unambiguous way.  
Recent angle resolved photoemission spectroscopy (ARPES) experiments 
\cite{ex6,ex8} have been able to shed new light on the symmetry of the 
OP, owing to their high resolution ($\sim$10\, meV) compared 
to the size of the superconducting gaps in the cuprates ($\sim$50\, meV). 
 For 
high quality Bi2212 single-crystal samples, the gap on the Fermi surface 
(FS) vanishes 
at two points symmetrically displaced from the 45$^o$ 
 direction in the Brillouin zone 
by about 10 degrees \cite{ex8} at the optimum dopant  
concentration ($\delta=0.17$)
   which has the highest $Tc$. Such a nodal structure is 
inconsistent with the $d_{x^2-y^2}$ symmetry, which has only one 
node per quadrant along the 45$^o$ direction. Also, a small admixture of 
an isotropic S-wave component to the 
$d_{x^2-y^2}$ OP does not increase the number of nodes but 
merely displaces the node away from the 45$^o$ direction. 

 Norman and coworkers 
have, in a recent set of papers \cite{fenor}, made a strong case 
for an OP whose symmetry is predominantly $S_{xy}$. They 
considered a phenomenological model of  electrons on a square 
lattice with an on-site repulsion ($V_0$) and attractive 
nearest neighbour ($V_1$) as well as next nearest neighbour ($V_2$) 
interactions. The Hamiltonian of the system is given by 
${\rm H} = {\rm H}_0 + {\rm H}_1$ where  

$$
 H_0 = \sum_{k,\sigma} (\epsilon_{\vec{k}} - \epsilon_F) 
      {c_{\vec{k},\sigma}^\dagger} {c_{\vec{k},\sigma}}\eqno(1a)\\ 
$$   
and 
$$  
  H_1  = V_0 \sum_{i} \hat{n}_{i,\uparrow} \hat{n}_{i,\downarrow} 
       + V_1 \sum_{i,\sigma,\sigma^\prime} \sum_{\delta_{nn}}
	\hat{n}_{i,\sigma} 
	\hat{n}_{i+\delta_{nn},\sigma^\prime}   
       + V_2 \sum_{i,\sigma,\sigma^\prime} \sum_{\delta_{nnn}}
	\hat{n}_{i,\sigma} 
       \hat{n}_{i+\delta_{nnn},\sigma^\prime}\eqno(1b)\\ 
$$  
The band dispersion $\epsilon_{\vec{k}}$ was chosen to fit the normal state 
quasiparticle dispersion obtained from the ARPES data on Bi2212 single
crystals \cite{fenor}. A six parameter 
tight binding Hamiltonian with [$t_0,...,t_5$] = 
[0.131,\,-0.149,\,0.041,-0.013,\,-0.014,\,0.013] (in eV) 
was found to reproduce the 
observed quasiparticle dispersion. Here $t_0$ is the   orbital energy,
 $t_1$ nearest 
neighbour (nn), $t_2$ next nearest neighbour (nnn) etc. hopping matrix 
elements.   
\addtocounter{equation}{1} 

A mean field  analysis of the spin-singlet pairing instabilities of 
the above model 
\cite{fenor}  showed that  the strongest 
instabilities correspond to OP's with $S_{xy}$ and 
$d_{x^2-y^2}$ symmetries because they best exploit 
the large single particle density of states (DOS) just below the 
FS; their
relative stability  is determined by the ratio of $V_1$ 
and $V_2$. A consistent fit to the ARPES data could be obtained in a range 
of values of $V_1$ and $V_2$ while $V_0$ was constrained to be  small. 

However, the above calculations  overlook the strong fluctuations 
that exist in the cuprates owing to their quasi two dimensional nature.
In this letter, we study the phase diagram of the Hamiltonian in Eq.(1) 
with the 
incorporation of the effects of fluctuations in the phase of the 
OP. We calculate the superfluid phase stiffness or helicity modulus
($\rho_s$) below the mean field transition temperature ($T_c^{MF}$), and use it in 
conjunction with the Kosterlitz-Thouless (KT) relation 
$\rho_s(T_c^{KT}) = {2\over\pi}k_BT_c^{KT}$ to determine the reduced 
transition temperature ($T_c^{KT}$). The interplanar Josephson coupling, 
that exists in the cuprates, does not lead to a strict KT like scenario 
as is being assumed here (except in  thin films). 
However, the magnitude of $T_c^{KT}$ 
is determined by $\rho_s$
and can therefore be expected to yield a good estimate of $T_c$ degradation 
due to phase fluctuations. With this improved bound on $T_c$, we determine 
the relative stability of superconducting 
OP's with B$_1$ ($d_{x^2-y^2}$), B$_2$ ($d_{xy}$) and A$_1$ (a 3 dimensional 
representation comprising $S,\,S^\star~{\rm and}~S_{xy}$) symmetries.
 We also study the nodal 
structure of the gap function on the FS. 

We find that the region of stability of the A$_1$ ($S_{xy}$) 
OP with respect to $d_{x^2-y^2}$ and $d_{xy}$, is reduced compared to the 
 mean field result. An OP 
which is predominantly $S_{xy}$ with very small 
isotropic S-wave ($S$) and extended S-wave ($S^\star$) components correctly 
reproduces the gap structure predicted from ARPES measurements. The best fit, 
corresponding to a $T_c^{KT}\sim 100 {\rm K}$ and a stable A$_1$ phase with 
a  nodal structure consistent with ARPES measurements, occurs for  
$V_0 \leq 15\,meV$, $V_1$ in the range 
$30-45\,meV$ and $V_2$ around   
$70-75\,meV$. For these parameters we find 
the variation of the transition temperature ($T_c^{KT}$) to be
qualitatively correct while
the angular positions of the nodes  on the FS
are nearly independent of  $\delta$.    

Standard  mean field factorization of the Hamiltonian (Eq.(1))
gives the gap equation 
\beq
\Delta_{\vec{k}}={1\over N} \sum_{\vec{k}^\prime} V(\vec{k}-\vec{k}^\prime) 
		  {{\Delta_{\vec{k}^\prime}}\over {2E_{\vec{k}^\prime}}} 
		  \tanh \left({{\beta E_{\vec{k}^\prime}}\over 2}\right) 
\enq 
where the quasiparticle energy is $E_{\vec{k}} = 
\sqrt{(\epsilon_{\vec{k}}-\epsilon_F)^2 + |\Delta_{\vec{k}}|^2}$ and  
$\Delta_{\vec{k}}$ is the BCS gap function. 
The pairing interaction, 
$V(q) = V_0 + 4 V_1(\cos q_x + \cos q_y) + 8 V_2 \cos q_x \cos q_y$,  
is separable and can be written as 
$V(\vec{k} - \vec{k}^\prime) 
      = \sum_{i=0}^4 \tilde{V}_i \eta_i(\vec{k}) \eta_i(\vec{k}^\prime) $
where $\eta_0(\vec{k}) = 1$, 
$\eta_1(\vec{k}) = \frac{1}{2}(\cos k_x + \cos k_y)$, 
$\eta_2(\vec{k}) = \frac{1}{2}(\cos k_x - \cos k_y)$, 
$\eta_3(\vec{k}) = \cos k_x\,\cos k_y$, 
$\eta_4(\vec{k}) = \sin k_x\,\sin k_y$, corresponding to 
[$S,\, S^\star,\, d_{x^2-y^2},\, S_{xy},\, d_{xy}$] symmetries, and 
$(\tilde{V}_0...\tilde{V}_4) = (V_0, 8V_1, 8V_1, 8V_2$, $8V_2)$. 
We have ignored terms corresponding to triplet pairing. It is easy 
to see that upon writing the gap functions as 
$\Delta_{\vec{k}} = \sum_i \eta_i(\vec{k}) \Delta_i$, the linearized 
gap equation (whose non-vanishing solution first appears at  
$T_c^{MF}$) factorizes into two independent equations for 
$\Delta_2$  and $\Delta_4$ (B$_1$ and B$_2$ representations) 
and three coupled linear equations involving 
$\Delta_0,\,\Delta_1,\,\Delta_3$ 
(A$_1$ representation). Since $\Delta_3$ is the largest component 
within the A$_1$ representation, the 
OP  symmetry  
is predominantly $S_{xy}$. We ignore the possibility of further transitions 
at lower temperatures. 

We now proceed to the calculation of the 
$\rho_s$ and $T_c^{KT}$.
We first solve the gap equations 
below $T_c^{MF}$, choosing the gap function 
$\Delta_{\vec{k}} = \sum_{i\in \cal{R}} \Delta_i \eta_i(\vec{k})$.
 As an illustration, we plot the gap functions  
$\Delta_0$, $\Delta_1$, $\Delta_3$ (A$_1$) and 
$\Delta_2$, $\Delta_4$ (B$_1$ and B$_2$ respectively) in the inset of 
Fig.1, as a function of temperature for optimal parameters (see below). 
To calculate $\rho_s$, we introduce 
a transverse vector potential  
and choose the gauge $A_y = 0$.  Then, the hopping matrix elements
 in $H_0$ (Eq.(1a)) acquire an additional phase factor according to the 
 Peierls substitution $t_{ij}\rightarrow t_{ij}\exp[{{{ie}\over{\hbar
c}}\int_{{\vec R}_j}^{{\vec R}_i}{\vec
A}\cdot {\vec dl}}]$.

The electron current operator ${\hat{j}}_x(\vec{R}_i)$ 
 can be found (to linear order in $A_x$) by 
differentiating $H_0$ with respect to $A_x(\vec{R}_i)$ and is of the form 
$$
{\hat{j}}_x(\vec{R}_i) = -c {{\partial H_0}\over{\partial A_x(\vec{R}_i)}}  
     = {\hat{j}}_x^{para}(\vec{R}_i) + {\hat{j}}_x^{dia}(\vec{R}_i)\eqno(3)
$$ 
The first term on the RHS of Eq.(3) is of zeroth order in $A_x$
and is   the electron velocity operator. It 
represents the paramagnetic
response of the quasi-particles.
The second term is of linear order in $A_x$ and
its physical origin is related to the  Meissner screening of the condensate.
The average diamagnetic current is obtained to be  
$$ 
j_x^{dia}(\vec{q}) = -{{e^2}\over{\hbar^2 c}} {1\over N} 
   \sum_{\vec{k},\sigma} \left\langle 
   c_{\vec{k},\sigma}^\dagger  c_{\vec{k},\sigma} \right\rangle 
   {{\partial^2\epsilon_{\vec{k}}}\over{\partial k_x^2}} 
   A_x(\vec{q})\eqno(4)                                          
$$ 
where the angular brackets correspond to an average in the mean field 
superconducting state.  From Eq.(4) we see that the diamagnetic
contribution to the
 phase stiffness ($\rho_s^{dia}$) is proportional to 
${1\over N}\sum_{\vec{k},\sigma}\left\langle 
c_{\vec{k},\sigma}^\dagger  c_{\vec{k},\sigma} \right\rangle 
{{\partial^2\epsilon_{\vec{k}}}\over{\partial k_x^2}} $,  the mean 
electronic kinetic energy along the x-direction \cite{scala}. 
Unlike models in 
which the electron dispersion is parabolic, $\rho_s^{dia}$ for  a lattice 
superconductor is not proportional to the average electronic density. 
Thus the use \cite{Carbotte} of the continuum expression 
$\rho_s^{dia}\propto -\sum_{\vec{k},\sigma} 
\left({{\partial\epsilon_{\vec{k}}}\over{\partial k_x}}\right)^2 {{\partial 
f(\epsilon _{\vec{k}}-\epsilon_{F})}\over{\partial \epsilon_{\vec{k}}}}$ 
(where $f$ is the Fermi function) is incorrect and could leave out 
important physics. 
\addtocounter{equation}{2} 

From linear response theory, the long wavelength paramagnetic current 
is found to be 
$$ 
j_x^{para}(\vec{q}) = - {1\over c} \left[\lim_{\vec{q}\to0} \lim_{\omega\to0}
     K^{xx}(\vec{q},\omega)\right]A_x(\vec{q})\eqno(5a) 
$$ 
where 
$K^{xx}(\vec{q},\omega) = -\imath \int dt\, \th(t)\,
    e^{\imath\omega t} \left\langle\left[ 
    j_x^{para}(\vec{q},t),\,j_x^{para}(-\vec{q},0)
    \right]\right\rangle $. 
We work in a transverse gauge and vertex corrections required to get
a gauge invariant current have been ignored.
The correlation function in Eq.(5a) is easily evaluated and found 
to be 
$$
\lim_{{\vec{q}}\to0} \lim_{\omega\to0} K^{xx}(\vec{q},\omega) 
    = {{e^2}\over{\hbar^2}} {1\over N} \sum_{\vec{k}} 
    \left({{\partial\epsilon_{\vec{k}}}\over{\partial k_x}}\right)^2 
    {{\partial f(E_{\vec{k}})}\over{\partial E_{\vec{k}}}}\eqno(5b) 
$$
\addtocounter{equation}{1} 
From Eqs (4) and (5) we obtain the superfluid phase stiffness 
\beq 
\rho_s(T) = {1\over{2N}} \sum_{\vec{k}} \left[ 
  \left({{\partial\epsilon_{\vec{k}}}\over{\partial k_x}}\right)^2 
  {{\partial f(E_{\vec{k}})}\over{\partial E_{\vec{k}}}}  + {1\over 2} 
  {{\partial^2\epsilon_{\vec{k}}}\over{\partial k_x^2}} 
  \left\{ 1- {{\epsilon_{\vec{k}}-\epsilon_F}\over{E_{\vec{k}}}} 
  \tanh \left({\beta E_{\vec{k}}\over 2}\right) \right\} \right]. 
\enq 

In Fig.1 we  show  $\rho_s$ as a function 
of temperature for  various OP symmetries. The results 
indicate that the temperature dependence of $\rho_s$ does 
not show any major qualitative change for the various types of OP's
 considered. Thus the  fluctuation corrected phase 
boundaries  are not expected to be 
very different from  the mean field case.

The KT transition temperature ($T_c^{KT}$) is found by solving  
\beq 
\rho_s(T_c^{KT}) = {2\over\pi}T_c^{KT}
\enq 
We use the  expression for $\rho_s$ (Eq.(6) in Eq.(7)), assuming mean field
values of the OP, and obtain $T_c^{KT}$.  
This procedure was employed earlier
by Denteneer et. al. \cite{Denteneer} in a study of the 2-d attractive 
 Hubbard model and  gives a qualitatively correct 
picture even in the limit of strong coupling where $T_c^{KT}$ is inversely 
proportional to the coupling. 
Since we neglect the renormalization of the OP (and hence 
$\rho_s$), due to the presence of vortex like fluctuations, the discontinous 
drop in $\rho_s$,  characteristic of KT transitions,
is not obtainable here. Hence our results 
give an upper bound on the true KT transition temperature. Another possible 
difficulty is that the expression in Eq.(7) has been 
derived for isotropic S-wave superconductors. However, even in 
the anisotropic case, the asymptotic form of the vortex-vortex 
interaction (at high vortex density) is logarithmic \cite{yren} and so 
corrections to Eq.(7) are likely to be small. A comparison of 
$T_c^{KT}$ thus obtained for different
OP's enable us to determine the phase diagram of the model
Hamiltonian. 

We now  discuss our results. For the sake of clarity 
we  present the phase diagrams in the $V_0-V_2$ and $V_1-V_2$ 
planes only. 
In Fig.2  we have drawn the phase diagram in the $V_0-V_2$ plane 
(we set $V_1=0$) for several values of   
$\delta$. For comparison, 
we also include the mean field phase boundaries \cite{fenor}. The dopant 
concentration ($\delta$) indicates the deviation from  half-filling. 
It is clear from Fig.2 that 
fluctuation effects are negligible at weak coupling ($|V_2|<50\, meV$). 
To obtain a stable $S_{xy}$ (A$_1$) solution at low filling, $V_0$ is 
required to be rather small ($V_0 < 0.4$ eV at $\delta = 0.17$), even 
smaller than the value indicated by the mean field analysis and 
larger values of $|V_2|/V_0$ are required 
to stabilize the $S_{xy}$ state. At high values of  $\delta$,
the A$_1$ phase is stable for much larger values of $V_0$ (a few eV) provided 
$|V_2|\leq 90 \, meV$. This is understandable as the onsite repulsion, which 
disfavours the A$_1$ state, has the largest effect 
close to half filling and becomes relatively ineffective at low
filling.

In Fig.3 we plot the phase boundary separating the $d_{x^2-y^2}$ and 
A$_1$ superconducting states in the $V_1-V_2$ plane (setting $V_0=0$). 
Fluctuation effects are once again seen to be negligible at weak coupling 
but lead to a larger critical value of $V_2/V_1$ (to stabilize the 
$S_{xy}$ state) at intermediate and strong couplings. Both the mean field and KT 
phase boundaries are  insensitive to $\delta$. 
This is not surprising as both the $d_{x^2-y^2}$ and $S_{xy}$ 
solutions have a very similar pairing density of states \cite{fenor} 
and thus the phase boundary is determined almost entirely by the 
interaction strength. 

We have found the values of interaction parameters at 
$\delta = 0.17$ (optimal doping) that produce a stable superconducting 
state with the A$_1$ symmetry and a transition temperature 
$T_c^{KT} \sim 100\,{\rm K}$  whose OP 
has the observed nodal structure. We find 
that the on-site repulsion $V_0$ is constrained to be  small 
($\leq 15\, meV$), $V_2$ lies in 
the range $70<- V_2 < 75\, meV$ whereas $V_1$ can be chosen in a 
relatively broader range $30< -V_1 < 45\, meV$. In Fig.4 we have plotted 
the angular dependence of $|\Delta_{\bf k}(T=0)|$ on the FS 
($\phi$ measures the
angular deviation  from the line passing through the $(0,\pi)$ and
$(\pi,\pi)$ points), for typical values of $V_0,\, V_1$ and $V_2$, lying 
within this range.  
 A puzzling feature of these results, which does not have 
any apparent microscopic justification, is the extremely small 
value of $V_0$, whereas the actual bare Cu onsite repulsion is known to be 
about a few $eV$.  

We have studied the variation of  $T_c^{KT}$ with 
 $\delta$ for the optimal parameters determined above. 
 We find that $T_c^{KT}$ has the correct
qualitative behaviour (see inset of Fig.5) with the highest $T_c^{KT}$ being 
achieved for $\delta\approx 0.25$. This is not surprising as the 
pairing density of states (for a superconducting instability in the 
A$_1$ state) at the Fermi energy is  peaked close to this value of 
$\delta$ \cite{fenor}.  
 We have also found the nodal structure of the OP 
as a function of $\delta$. The magnitude of the gap function on 
the FS away from the nodes is strongly dependent on 
$\delta$ (reflecting the variation in $T_c^{KT}$ ). However the 
angular positions of the nodes on the FS
 are insensitive to  $\delta$ (see 
 Fig.5 and its inset) and stay close to their positions for $\delta=0.17$. 

In conclusion, we have studied 
the competition between superconducting OP's 
with different symmetries within a 
phenomenological BCS model and obtained a 
phase diagram which incorporates the effects of  fluctuations. Our 
phase diagram, together with the 
magnitude of the transition temperature and the nodal structure of the 
OP, enable us to identify the best choice of parameters appropriate to the 
optimally doped Bi2212 compound.  We find that the transition
temperature has the correct doping dependence and the angular positions of
the nodes of the gap function on the FS are nearly independent
of $\delta$ for the best parameters.

\newpage 

\noindent {\bf FIGURE CAPTIONS}

\noindent Fig.1.  $\rho_s$ as a function of temperature at $\delta=0.17$, 
$V_0=15\,meV,\,\, V_1=-40\,meV$ and $V_2=-75\,meV$ for the A$_1$, 
B$_1$ and B$_2$ symmetries. The intersection of the straight line
with $\rho_s$ gives $T_c^{KT}$. Inset shows the variation of the
mean field OP's with temperature.

\noindent Fig.2.  The mean field (dashed lines) and KT (solid lines) 
phase boundaries indicating the regions of stability of the A$_1$ and
d$_{xy}$ phases are shown in the ($V_0,-V_2$) plane for various 
$\delta$. The inset shows the KT phase diagram for $\delta=0.17$
on an expanded scale.

\noindent Fig.3.  The mean field  and KT  phase
boundaries indicating the regions of stability of the A$_1$ and
d$_{x^2-y^2}$ phases are shown in the ($-V_2,-V_1$) plane for various 
$\delta$. The inset shows the KT phase diagram for $\delta=0.17$
on an expanded scale.

\noindent Fig.4. The angular dependence of the gap function $|\Delta_{\bf
k}(T=0)|$ on the Fermi surface is shown for $\delta=0.17$ and various
values of interaction parameters in the optimal range. The angles are
measured with respect to the $(\pi,\pi)$ point as in ref. [10].

\noindent Fig.5. $|\Delta_{\bf k}(T=0)|$ on the Fermi surface with
$V_0=15\,meV\,$, 
$V_1=-40\,meV$ and $V_2=-74\,meV$ (optimal parameters)
for various $\delta$. Curves 1-5 correspond to 
$\delta$= 0.10,\,0.13,\,0.17,\,0.21 and 0.26  respectively. Inset shows the 
angular position ($\alpha$) of the node on the Fermi surface 
(measured from the $45^o$ direction) and $T_c^{KT}$ (in $^o $K)
as a function of $\delta$ for optimal parameters.
	 
\end{document}